\documentclass[twocolumn,aps,superscriptaddress,showpacs,floatfix]{revtex4-2}
\usepackage{graphicx}
\usepackage{dcolumn}
\usepackage{bm}
\usepackage[colorlinks,linkcolor=blue,anchorcolor=blue,citecolor=blue,urlcolor=blue]{hyperref}
\usepackage{amsmath}
\usepackage{multirow}
\usepackage{textcomp}
\usepackage{xcolor}
\usepackage{soul}
\usepackage{booktabs}
\usepackage{tabularx}
\usepackage{multirow}
\usepackage{threeparttable}
\usepackage{CJKutf8}
\begin{document}
\begin{CJK*}{UTF8}{gbsn}
\preprint{APS/123-QED}
\title{Best Reaction Target To Determine Proton Distribution Radii of Atomic Nuclei}

\author{Jun-Yao Xu(徐俊瑶)}
\affiliation{School of Physics, Beihang University, 100191 Beijing, People's Republic of China}

\author{Bao-Hua Sun(孙保华)}
\email{bhsun@buaa.edu.cn}
\affiliation{School of Physics, Beihang University, 100191 Beijing, People's Republic of China}

\author{Isao Tanihata(\CJKfamily{min}谷畑勇夫)} 
\email{tanihata@rcnp.osaka-u.ac.jp}
\affiliation{School of Physics, Beihang University, 100191 Beijing, People's Republic of China}

\author{Satoru Terashima(\CJKfamily{min}寺嶋知)}
\affiliation{School of Physics, Beihang University, 100191 Beijing, People's Republic of China}
\affiliation{Institute of Modern Physics, Chinese Academy of Sciences, Lanzhou 730000, China}

\author{Jian-Wei Zhao(赵建伟)}
\email{zhaojianwei@buaa.edu.cn}
\affiliation{School of Physics, Beihang University, 100191 Beijing, People's Republic of China}

\author{Ji-Chao Zhang(张寂潮)} 
\affiliation{School of Physics, Beihang University, 100191 Beijing, People's Republic of China}

\author{Ge Guo(郭戈)} 
\affiliation{School of Physics, Beihang University, 100191 Beijing, People's Republic of China}

\author{Shi-Tao Wang(王世陶)} 
\affiliation{Institute of Modern Physics, Chinese Academy of Sciences, Lanzhou 730000, China}

\author{Lei Shen(沈雷)} 
\affiliation{Shanghai Institute of Applied Physics, Chinese Academy of Sciences, Shanghai 201800, China}

\author{Jun Su(苏军)}
\affiliation{Sino-French Institute of Nuclear Engineering and Technology, Sun Yat-sen University, Zhuhai 519082, China}

\author{Xiao-Dong Xu(徐晓栋)} 
\affiliation{Institute of Modern Physics, Chinese Academy of Sciences, Lanzhou 730000, China}

\author{Andrej Prochazka}
\affiliation{Medaustron, 2700 Wiener Neustadt, Austria}

\author{Guang-Shuai Li(李光帅)} 
\affiliation{School of Physics, Beihang University, 100191 Beijing, People's Republic of China}

\author{Xiu-Lin Wei(魏秀琳)} 
\affiliation{School of Physics, Beihang University, 100191 Beijing, People's Republic of China}

\author{Chang-Jian Wang(王长建)} 
\affiliation{School of Physics, Beihang University, 100191 Beijing, People's Republic of China}

\author{Feng Wang(王枫)} 
\affiliation{School of Physics, Beihang University, 100191 Beijing, People's Republic of China}

\author{Meng Wang(王萌)} 
\affiliation{School of Physics, Beihang University, 100191 Beijing, People's Republic of China}

\author{Jing Wang(王晶)} 
\affiliation{School of Physics, Beihang University, 100191 Beijing, People's Republic of China}

\author{Liu-Chun He(何鎏春)} 
\affiliation{School of Physics, Beihang University, 100191 Beijing, People's Republic of China}

\author{Chuan-Ye Liu(刘传业)} 
\affiliation{School of Physics, Beihang University, 100191 Beijing, People's Republic of China}

\author{Wen-Jian Lin(林文健)} 
\affiliation{School of Physics, Beihang University, 100191 Beijing, People's Republic of China}

\author{Wei-Ping Lin(林炜平)} 
\affiliation{Key Laboratory of Radiation Physics and Technology of the Ministry of Education, Institute of Nuclear Science and Technology, Sichuan University, Chengdu 610064, China}

\author{Zhong Liu(刘忠)} 
\affiliation{Institute of Modern Physics, Chinese Academy of Sciences, Lanzhou 730000, China}
\affiliation{School of Nuclear Science and Technology, University of Chinese Academy of Sciences, Beijing 100049, China}

\author{Pei-Pei Ren(任培培)} 
\affiliation{Key Laboratory of Radiation Physics and Technology of the Ministry of Education, Institute of Nuclear Science and Technology, Sichuan University, Chengdu 610064, China}

\author{Yu Zhang(张宇)} 
\affiliation{School of Physics, Beihang University, 100191 Beijing, People's Republic of China}

\author{Mei-Xue Zhang(张湄雪)} 
\affiliation{School of Physics, Beihang University, 100191 Beijing, People's Republic of China}

\author{Ya-Zhou Sun(孙亚洲)} 
\affiliation{Institute of Modern Physics, Chinese Academy of Sciences, Lanzhou 730000, China}

\author{Zhi-Yu Sun(孙志宇)} 
\affiliation{Institute of Modern Physics, Chinese Academy of Sciences, Lanzhou 730000, China}

\author{Chen-Gui Lu(鲁辰桂)} 
\affiliation{Institute of Modern Physics, Chinese Academy of Sciences, Lanzhou 730000, China}

\author{Xue-Heng Zhang(章学恒)} 
\affiliation{Institute of Modern Physics, Chinese Academy of Sciences, Lanzhou 730000, China}

\author{Jin-Rong Liu(刘锦蓉)} 
\affiliation{School of Physics, Beihang University, 100191 Beijing, People's Republic of China}

\author{Tian-Yu Wu(吴天宇)} 
\affiliation{School of Physics, Beihang University, 100191 Beijing, People's Republic of China}

\begin{abstract}
We found that a heavy target such as Pb is most suitable for determining the proton distribution radii of unstable nuclei through charge-changing cross-section ($\sigma_\text{cc}$) measurements. As a heavy ion probe, low-$Z$ targets are routinely used to determine nucleon distribution radii of unstable isotopes. This approach has recently been extended to study proton distribution radii from $\sigma_\text{cc}$ measurements. However, empirical scaling factors have to be introduced to apply the Glauber models. In the present work, we systematically investigated the scaling factor using 39 new $\sigma_\text{cc}$ data of 18 $p$-shell nuclei on hydrogen, carbon, silver, and lead targets at around 240 MeV/nucleon. Together with the existing data, we reveal a universal dependence of the scaling factor on both the masses of target nuclei and the separation energies of projectile nuclei. The scaling factors decrease with increasing target-nucleus mass and converge to 1 for the highest-$Z$ target, making the scaling unnecessary. We conclude that instead of a low-$Z$ target, employing a heavy target such as Pb in $\sigma_\text{cc}$ measurements is the best option to determine the proton distribution radii of unstable nuclei.

\end{abstract}

\maketitle
\end{CJK*}

{\it Introduction}---The nucleon, neutron, and proton distribution radii are fundamental to understanding how protons and neutrons are bound in atomic nuclei. 
In what follows, we denote the nucleon, neutron, and proton distribution radii as the matter radius, neutron radius, and proton radius ($R_p$) of a nucleus, respectively. Separate determinations of neutron and proton radii provide information on the neutron (or proton) skin of nuclei, which is crucial for revealing the structure of exotic nuclei, constraining the equation of state of asymmetric nuclear matter and neutron stars, and examining state-of-the-art nuclear models.  

High-energy heavy-ion collisions have been used to determine the matter radii of unstable nuclei since the 1980s~\cite{Tanihata1985PRL}. 
The analysis of interaction cross sections ($\sigma_\text{I}$) on low-$Z$ targets with the Glauber model provides a means to determine matter radii of numerous unstable nuclei~\cite{OZAWA2001NPA}, including those far from the stability line~\cite{Bagchi2020PRL,Tanaka2020PRL}. 
$\sigma_\text{I}$ is insensitive to the target mass for targets with $Z \ge 4$ (Be), and essentially no visible sensitivity was observed for Be, C, and Al targets~\cite{Tanihata1988PLB,Tanihata2013PPNP}. 
For targets with $Z$ higher than that of Al, $\sigma_\text{I}$ has not been used for radius determinations because the Coulomb interaction becomes significant. Instead, electromagnetic dissociation (EMD) has been employed to study neutron-halo nuclei~\cite{Kobay1989PLB,Nakamura2006PRL,Nakamura2009PRL,Cook2020PRL}. In these cases, the enhancement of EMD occurs primarily due to $E1$ photons from the target nuclei.
In particular, the removal of halo neutrons is enhanced because halo neutrons are separated from the core part of the nuclei, which contains all protons.

Probing the proton radii of unstable nuclei experimentally represents a central and highly challenging endeavor in nuclear physics. Recent collinear laser spectroscopy measurements have extended the accessible range of exotic isotopes~\cite{Gusta2025PRL}. The internal target technique offers a promising approach for electron scattering studies of unstable isotopes near the $\beta$-stability line~\cite{Tsukada2023PRL}.
As a complementary method to the above electromagnetic probes, 
charge-changing cross sections ($\sigma_{\text{cc}}$), aligned with the interaction cross sections, have been investigated for their correlation with $R_p$ in the Glauber model~\cite{MengJ2002PLB}. This method can be readily accomplished at the in-flight fragment separator equipped with standard beam-line detectors, particularly when coupled with the high-energy multi-step fragmentation process~\cite{Wei2025NST}, and can access unstable isotopes with lifetimes down to several hundred nanoseconds. Systematic data can be used to constrain the equation of state~\cite{Xujy2022PLB,Aumann2017PRL}.
Moreover, precision $\sigma_{\text{cc}}$ data are important inputs in fields like heavy-ion therapy, radiation protection~\cite{Durante2016RPIP} and galactic cosmic-ray propagation~\cite{YanQ2020NPA}. 

In the Glauber framework, only the projectile protons are assumed to participate in the collision, while the projectile neutrons are spectators. However, a systematic underestimation of $\sim 10\%$ by such a Glauber model calculation was revealed relative to experimental data~\cite{Yamag2010PRC}. 
Empirically, one has to introduce a scaling factor in Glauber model calculations to explain the $\sigma_{\text{cc}}$ data of stable isotopes, whose $R_p$ are known from electron scattering. 
This method, normalizing the scaling factor by stable nuclei, 
has been applied to determine the $R_p$ of many neutron-rich nuclei~\cite{Yamag2011PRL,Ozawa2014PRC,Estra2014PRL,Bagchi2019PLB}.  
To explain the systematic deviation in calculating the $\sigma_{\text{cc}}$, the charged-particle evaporation (CPE) mechanism has been proposed ~\cite{Tanaka2022PRC,ZhaoJWPLB2023}. In this process, removing only neutron(s) from the projectile nucleus in collisions leaves the pre-fragment in unbound states above the charged-particle emission threshold. The pre-fragment subsequently de-excites via charged-particle emission (e.g., protons, deuterons, alpha particles), thereby contributing to the charge-changing cross section. 

Although several approaches have been developed to correct or incorporate this CPE channel~\cite {Estra2014PRL,Bagchi2019PLB,Tanaka2022PRC,ZhaoJWPLB2023}, they have not yet converged into a unified, thoroughly validated framework.
 Recently, a strong correlation was revealed between the scaling factor and the neutron and proton separation energies of the projectile nuclide ($S$-factor)~\cite{Zhang2024SB,zhao2024PLB}, allowing one to deduce $R_p$ in a unified way. Nearly identical $R_p$ values were deduced from $\sigma_{\text{cc}}$ measurements on both carbon and hydrogen targets for the neutron-rich carbon isotopes. However, $R_p$ values of neutron-rich nitrogen isotopes deduced from the hydrogen target are systematically smaller than those from the carbon target. This may reflect that proton or carbon targets can exhibit different sensitivity to the density distribution~\cite{Aumann2021PPNP}, although the $S$-factor correction can largely correct for the reaction mechanism, with one proton evaporation after one neutron removal.

As discussed, the scaling factor correlates with the particle separation energies of the nuclei of interest.  
The insight raises a central question regarding the influence of the reaction target on the extraction of $R_p$. 
To directly quantify this target dependence and identify a unified experimental approach, we performed new measurements of 39 $\sigma_\text{cc}$ of 18 nuclei, $^{9,10}$Be, $^{10-13}$B, $^{12-16}$C, $^{14-17}$N, and $^{16-18}$O, on hydrogen, carbon, silver, and lead targets with beam energy at around 240 MeV/nucleon. All are reported for the first time. Leveraging this consistent dataset, we observe a systematic trend in the scaling factor across different target nuclei and conclude that the highest-$Z$ target (i.e., Pb) is optimal for determining the proton radii of unstable nuclei, allowing us to extract charge radii of unstable isotopes with less model uncertainty.

\label{section2}

\begin{figure}[ht]
\includegraphics[width=0.5 \textwidth]{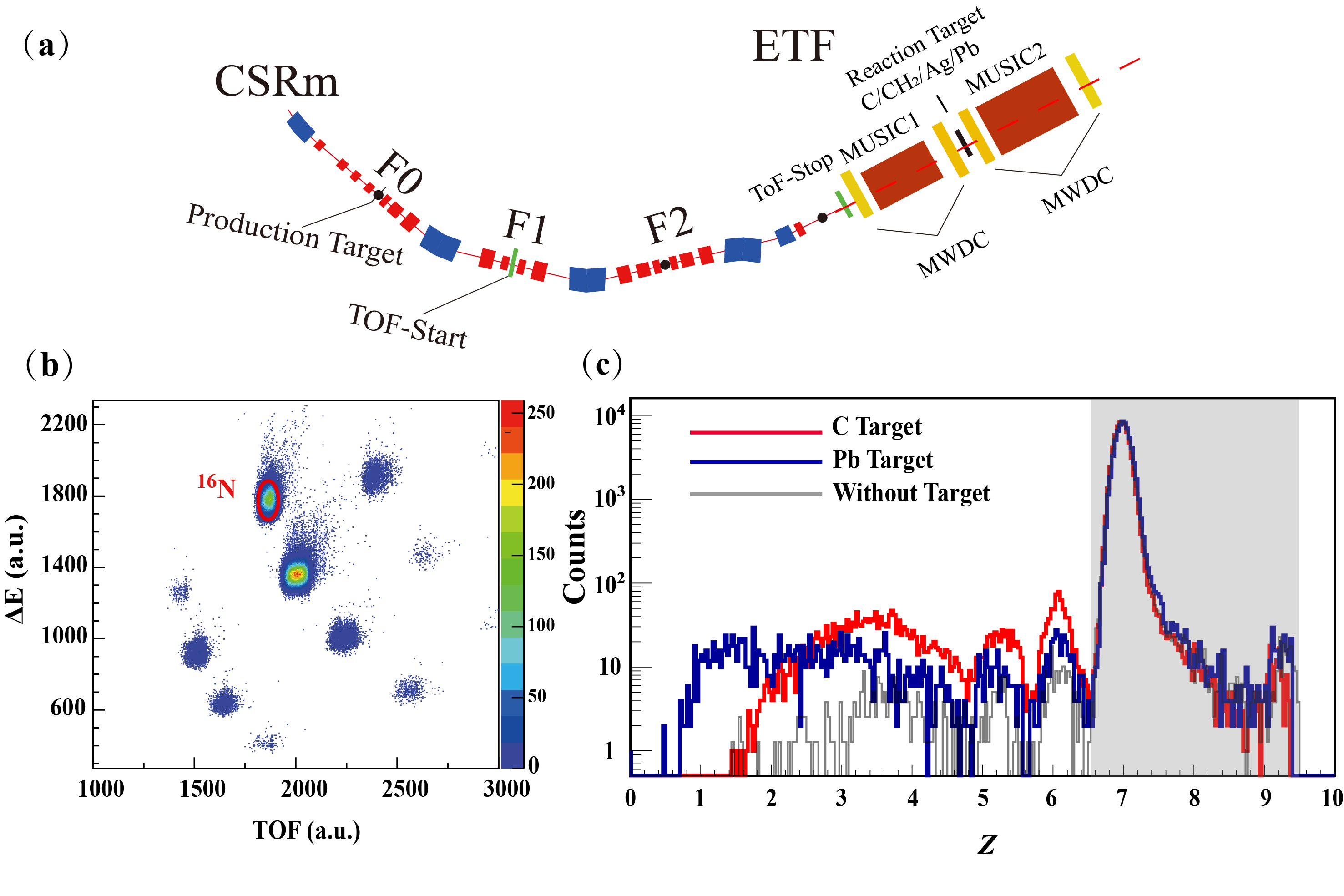}
\centering
\caption{(Color online) (a) Schematic view of the experimental facility including RIBLL2 and ETF. (b) Particle identification spectrum before the reaction target. (c) $Z$ identification spectrum after the
reaction target by tagging the incident ions $^{16}$N. The red, blue, and gray histograms correspond to the C, Pb, and empty-target data, respectively. The $Z=7$ peak heights for the Pb and empty targets are normalized relative to those from the C target. The shaded area represents the $N_\text{out}$ selection window.}
\label{fig:setup}
\end{figure}

{\it Experiment}---The experiment was conducted at the Heavy Ion Research Facility in Lanzhou (HIRFL)~\cite{ZhanWL2008NPA}.
A schematic of the experimental setup is shown in Fig.~\ref{fig:setup}(a). The isotopes of interest were produced by fragmenting a 280 MeV/nucleon $^{18}$O primary beam on a 15-mm-thick Be target. Secondary beams were separated and identified event-by-event in flight using the second Radioactive Ion Beam Line (RIBLL2)~\cite{BHSun2018SB},
based on measurements of their magnetic rigidity ($B\rho$), time-of-flight (TOF) between the F1 plane and the External Target Facility (ETF) with plastic scintillators, and energy loss ($\Delta E$) in a multi-sampling ionization chamber (MUSIC)~\cite{ZhangXH2015NIMA}.
A typical particle identification spectrum obtained before the reaction target is presented in Fig.~\ref{fig:setup}(b).
The secondary beams then impinged on the reaction targets of carbon (0.8999 g/cm$^2$), polyethylene (0.9557 g/cm$^2$), natural silver (2.0943 g/cm$^2$), and lead (2.2844 g/cm$^2$), respectively.
An empty target frame was employed to evaluate background contributions. The beam energy was degraded to approximately 240 MeV/nucleon at the target center. Details regarding the experimental setup and detector performance can be found in Refs.~\cite{Wangcj2023CPC,LiGS2023PRC}.

$\sigma_\text{cc}$ is measured using the transmission method by counting the numbers of incident nuclei ($N_\text{in}$) and outgoing particles without proton loss ($N_\text{out}$).
It is derived through the equation, $\sigma_\text{cc}=-{t}^{-1}\ln({\Gamma}/{\Gamma_0})$,
where $t$ is the target thickness, and $\Gamma$ and $\Gamma_0$ are the ratios $N_\text{out}/N_\text{in}$ with and without the reaction target, respectively. 
Taking the case of $^{16}$N shown in Fig.~\ref{fig:setup}(b) as an example, 
the corresponding charge identification of outgoing particles was done using the MUSIC2 detector. 
To obtain the number of $N_{\text{out}}$, a selection window was placed on the $Z \ge 7$ region as indicated by the shaded area in Fig.~\ref{fig:setup}(c). The lower boundary was set to $4\sigma$ below the centroid of the $Z = 7$ peak, with $\sigma$ determined from a Gaussian fit.
There is a tail in the higher-atomic-number region, but this effect can be completely removed by analyzing the empty-target data.

\label{section3}
{\it Results and discussion}---The $\sigma_\text{cc}$ measured in this work are summarized in Table~\ref{tab:cccs_data}. Cross sections on hydrogen are obtained by subtracting the cross sections on carbon from those on the polyethylene target.
Our new results for $^{11}$B on hydrogen, $^{10}$Be and $^{12,14-16}$C on carbon, and $^{12}$C, $^{14}$N, and $^{16}$O on lead are well consistent with Refs.~\cite{Webber1998Aj,Yamag2011PRL,Zeitlin2007PRC, Zeitlin2011PRC}. 
For comparison, we performed calculations using both the zero-range and finite-range~\cite{AbuIb2008PRC} optical-limit Glauber models (GM). The calculations include Coulomb corrections to the trajectory, which are essential for high-$Z$ targets~\cite{Charagi1990PhysRevC}. The nuclear charge radii for the relevant nuclei are well known~\cite {zhao2024PLB,Zhang2024SB,Angli2013ADND,auser2009PRL,Estra2014PRL}, from which one can deduce $R_p$ and the harmonic oscillator (HO) density distributions. 
The target densities were modeled using a Dirac delta function for $^{1}$H, an HO distribution for $^{12}$C, and a two-parameter Fermi distribution for natural Ag and Pb.
The finite-range GM results exceed the zero-range results by 0.6--2.6\%, but both calculations systematically underestimate the $\sigma_\text{cc}$ measured on carbon and hydrogen.
Subsequent analysis shows that finite- and zero-range GM results lead to the same conclusion. Hereafter, we present only zero-range GM predictions.

\begin{table}[h]
    \caption{The charge-changing cross sections measured in the present work on H ($\sigma_\text{cc,H}^\text{exp}$), C ($\sigma_\text{cc,C}^\text{exp}$), $^\text{nat}$Ag ($\sigma_\text{cc,Ag}^\text{exp}$), and $^\text{nat}$Pb ($\sigma_\text{cc,Pb}^\text{exp}$) targets.}
    \label{tab:cccs_data}
    \centering
    \setlength\tabcolsep{1.1pt}
    \renewcommand{\arraystretch}{1}
    \begin{threeparttable}
    \begin{tabular}{cccccc}
        \hline
        &Energy&$\sigma^\text{exp}_\text{cc, H}$& $\sigma^\text{exp}_\text{cc, C}$& $\sigma^\text{exp}_\text{cc, Ag}$&$\sigma^\text{exp}_\text{cc, Pb}$\\
        &(MeV/nucleon)&(mb)&(mb)&(mb)&(mb)\\
        \hline
        $^{9}$Be	&	244(10)	&	280(81)	&	565(94)	&	1867(452)	&	2841(774)		\\
        $^{10}$Be	&	240(3)	&	170(44)	&	645(61)	&	2244(296)	&	3095(384)	\\
        $^{10}$B	&	234(11)	&	 	    &	572(74)	&	 	        &	 	        \\
        $^{11}$B	&	234(10)	&	130(50)	&	609(52)	&	1793(271)	&	2903(500)	\\
        $^{12}$B	&	234(8)	&	117(34)	&	740(43)	&	2478(240)	&	3147(282)	\\
        $^{13}$B	&	244(7)	&	154(19)	&	710(25)	&	2375(99)	&	2946(166)	\\
        $^{12}$C	&	229(9)	&	 	    &	728(22)\tnote{1}	    &	 &	2506(358)	\\
        $^{13}$C	&	232(8)	&	 	    &	720(25)\tnote{1}	    &	 &	 	\\
        $^{14}$C	&	235(7)	&	154(11)	&	703(12)\tnote{1}	&	2076(66)	&	2911(95)	\\
        $^{15}$C	&	237(7)	&	171(16)	&	744(20)\tnote{1}	&	2272(83)	&	2773(127)	\\
        $^{16}$C	&	238(6)	&	178(16)	&	744(19)\tnote{1}	&	2371(74)	&	3134(129)	\\
        $^{14}$N	&	225(7)	&	 	    &	839(31)\tnote{1}	&	 	&	3249(511)	\\
        $^{15}$N	&	228(6)	&	 	    &	808(15)\tnote{1}	&	 	&		\\
        $^{16}$N	&	233(4)	&	208(18)	&	859(22)\tnote{1}	&	2511(110)	&	3014(164)	\\
        $^{17}$N	&	237(3)	&	191(14)	&	834(20)\tnote{1}	&	2312(71)	&	3132(110)	\\
        $^{16}$O	&	221(5)	&		    &	863(18)\tnote{1}	&	 	&	2826(246)	\\
        $^{17}$O	&	227(4)	&		    &	869(80)\tnote{1}	&	 	&	 	\\
        $^{18}$O	&	259	    &		    &	848(11)\tnote{1}	&	 	&	 	\\
        \hline
    \bottomrule
    \end{tabular}
    \begin{tablenotes}
    \footnotesize
    \item[1] Data measured with the same setup, already published in 
Ref.~\cite{zhao2024PLB}
    \end{tablenotes}
    \end{threeparttable}
\end{table}

The underestimation by the GM is attributed to the fact that the calculation includes only the direct charged-particle removal process ($\sigma^\text{cal}_\text{cc}$), while neglecting contributions from indirect reaction channels ($\sigma^\text{indir}_\text{cc}$).
The indirect contributions primarily include charged-particle emission following pure neutron removal or inelastic scattering. The latter includes EMD and nuclear inelastic excitation. EMD is induced when the projectile nucleus passes by a high-$Z$ target at high speed, leading to the emission of charged particles from the projectile, with the ($\gamma$, $p$) channel being the dominant contributor to $\sigma_\text{cc}$~\cite{Bertu1988PR,LiuJR2025PRC}. At the present beam energy regime, nuclear inelastic excitation and its associated Coulomb--nuclear interference are negligibly small~\cite{Bertu1999PR}. Moreover, inelastic excitation is dominated by Coulomb rather than nuclear interaction for high-$Z$ targets~\cite{Aumann1999PRC}.
The experimental charge-changing cross section ($\sigma^\text{exp}_\text{cc}$) can be expressed as 
\begin{center}
  \begin{equation}
  \begin{split}
    \sigma^\text{exp}_\text{cc}&=\sigma^\text{cal}_\text{cc}+\sigma_\text{indir} \\
 & \cong\sigma^\text{cal}_\text{cc}+\sigma_\text{evap}+\sigma_\text{EMD}^p\;\\
                          & \cong\sigma^\text{cal}_\text{cc}+\sum_i{ P^\text{evap}_{i} \sigma_{-in}}+\int N(E_\gamma) \sigma_\gamma^p(E_\gamma)\frac{dE_\gamma}{E_\gamma}\;,
  \end{split}
\label{eq:cc}
\end{equation}
\end{center}
where $\sigma_\text{evap}$ and $\sigma_\text{EMD}^p$ are the cross sections of the CPE after removing only neutrons and electromagnetic dissociation, respectively.
$\sigma_{-in}$ is the cross section of removing only $i$ neutrons to form the pre-fragments. $P^\text{evap}_{i}$ refers to the corresponding CPE probability, which strongly depends on the excitation energy distribution (EED) of the corresponding neutron-removed pre-fragment. $N(E_\gamma)$ and $\sigma_\gamma^p$ represent the virtual-photon spectrum and the photonuclear proton-emission cross section, respectively.

\begin{figure}
\includegraphics[width=0.45\textwidth]{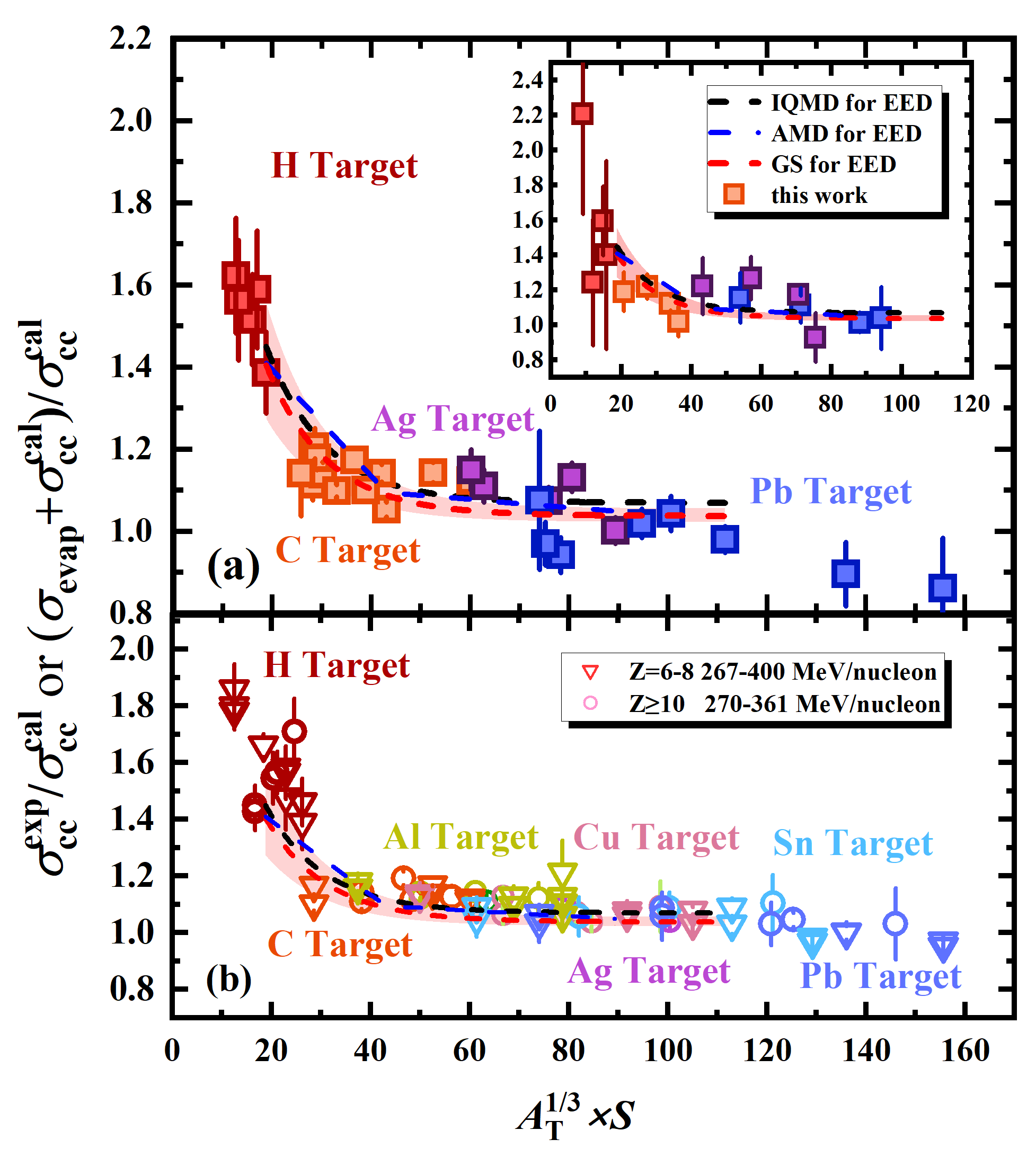}
\centering
\caption{(Color online) Ratios of experimental charge-changing cross sections to theoretical ones, $\sigma^\text{exp}_\text{cc}$/$\sigma^\text{cal}_\text{cc}$, as a function of $A_\text{T}^{1/3}\times S$. $S$ is in the unit of MeV.
The present data are shown as filled squares with Be and B isotopes in the inset in panel (a). The literature data, grouped into two $Z$ ranges, are shown as open symbols in panel (b). Various targets are distinguished in colors. 
The red, blue, and black dashed lines represent the calculated ratios ($\sigma_\text{evap}+\sigma_\text{cc}^{\text{cal}})/\sigma_\text{cc}^{\text{cal}}$ of the projectile $^{14}$C on various targets, with excitation energy distributions provided by the GS,  
AMD, and IQMD models, respectively. Refer to the text for details.  
}
\label{fig:ratio_target}
\end{figure}

Experimentally, the contribution of indirect reaction channels can be characterized by the 
ratios of experimental charge-changing cross sections to theoretical ones, $\sigma^\text{exp}_\text{cc}$/$\sigma^\text{cal}_\text{cc}$. A strong correlation is observed between the ratio and the nucleon separation energies, $S_n(A,Z)+S_p(A-1,Z)$, as well as a notable difference between carbon and proton targets~\cite{Zhang2024SB}. 
$S_n(A,Z)$ and $S_p(A-1,Z)$ are the one-neutron separation energy of the projectile nucleus ($A$, $Z$) and the one-proton separation energy of the nucleus produced by removing only one neutron from the projectile nuclide, respectively.  

To transform empirical observations into predictive power, we introduce a new quantity, $A_\text{T}^{1/3}\times S(A, Z)$, where $S(A, Z)= [S_n(A,Z)+\text{Min}(S_p(A-1,Z),S_\alpha(A-1,Z))]$.
It incorporates the target-size dependence and the competition between proton and alpha-particle emission during de-excitation, the two dominant de-excitation channels. Here, $A_\text{T}$ is the mass number of a target nuclide, and $S_\alpha(A-1,Z)$ is the one-alpha separation energy of the nucleus $(A-1, Z)$.
The minimum function, $\text{Min}(S_p(A-1, Z), S_\alpha(A-1, Z))$, selects the lower of the two separation energies, characterizing the preferred charged-particle evaporation channel
during the de-excitation of pre-fragments.

Fig.~\ref{fig:ratio_target} shows $\sigma^\text{exp}_\text{cc} / \sigma^\text{cal}_\text{cc}$ as a function of $A_\text{T}^{1/3}\times S$.  
Our data for C, N, and O isotopes show that the ratio decreases monotonically from $\sim$1.7 on hydrogen to $\sim$1 on lead as $A_\text{T}^{1/3}\times S$ increases from about 10 to 110 (fm$\cdot$ MeV) in Fig.~\ref{fig:ratio_target} (a). The same trend is observed for Be and B isotopes in the inset.  
Here, nuclei $^{9}$Be and $^{10}$B are excluded because their one-neutron removal is followed by breakup,  
invalidating a description based on separation energies.
To extend the investigation beyond the current $p$-shell data, we have surveyed existing measurements. 
There are 30 data points for $sd$-shell nuclei ($^{20}$Ne, $^{28}$Si, $^{36,40}$Ar)~\cite{Zeitlin2007NPA,Zeitlin2008PRC,Zeitlin2011PRC,IANCU2005RM} and 42 data points for $p$-shell nuclei~\cite{Webber1990PRC,Webber1998Aj,Schall1996NIM,Cecchini2008NPA,Zeitlin2007NPA,Zeitlin2007PRC,Zeitlin2011PRC}, with reaction energies spanning over 267--400 MeV/nucleon on hydrogen, carbon, aluminum, copper, silver, tin, and lead. These independent datasets in Fig.~\ref{fig:ratio_target} (b) lead to the same conclusions. The convergence at large $A_\text{T}^{1/3}\times S$ values points to a relative suppression of indirect reaction channels. Consequently, it enables extracting proton radii directly from $\sigma_\text{cc}^\text{exp}$, eliminating the need for any empirical scaling corrections to the Glauber model.
 
The following discussion addresses how the scaling factor reaches 1 for the higher-$Z$ target. We take $^{14}$C as an example and compute its evaporation contribution on various targets in the Abrasion-Ablation model. The EEDs, the key input for $P^\text{evap}_{i}$ in Eq.~\ref{eq:cc}, are calculated by the microscopic isospin-dependent quantum molecular dynamics (IQMD)~\cite{JunSu2011PRC}, antisymmetrized molecular dynamics (AMD)~\cite{Shen2024PRC} models, and a parameter-adjusted Gaimard-Schmidt (GS) model~\cite{JJGaimard1991NPA}. In the GS model, the maximum excitation energy ($E_\text{max}$) was set to 30$^{+16}_{-11}$ MeV to reproduce our experimental data for the carbon isotopes on carbon. The ratios of $(\sigma_\text{evap}+\sigma^\text{cal}_\text{cc})/\sigma^\text{cal}_\text{cc}$ are presented in Fig.~\ref{fig:ratio_target}. The overall trend of the calculations agrees well with the data within experimental uncertainties. 
We note that the $P^\text{evap}_{1}$ can differ by more than a factor of 3 in the three models; however, the ratios are much less sensitive to this variation.

\begin{figure}
\includegraphics[width=0.45\textwidth]{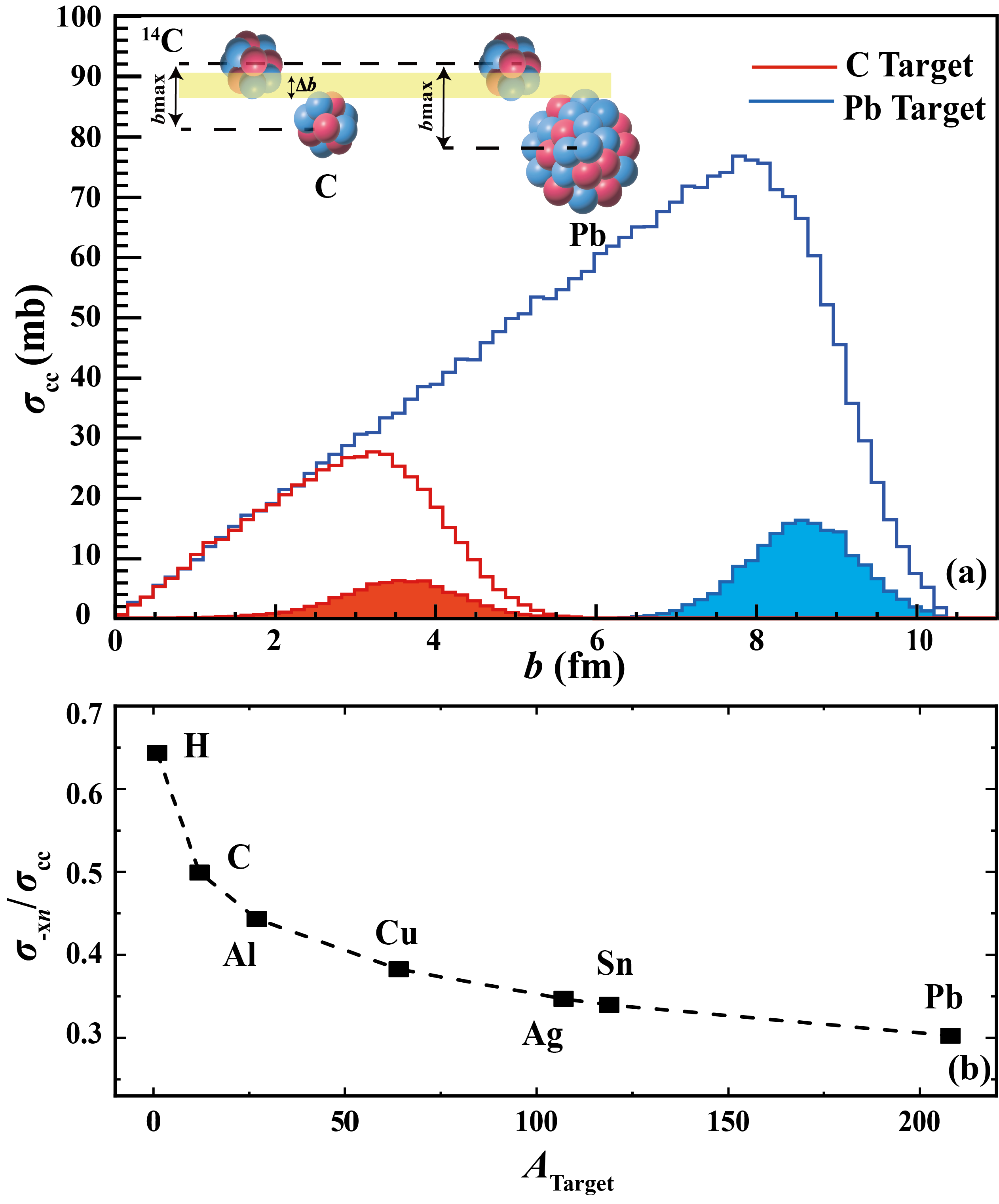}
\centering
\caption{(Color online)
(a) 
Charge-changing cross sections as a function of impact parameter for $^{14}\text{C}$ colliding on carbon (red) and lead (blue) in the IQMD+GEMINI approach. The bin width is $0.157\text{ fm}$. The open and filled histograms represent the total charge-changing cross sections and CPE, respectively.
(b) Target-size dependence of neutron removal fraction in the geometric model.
}
\label{fig:IQMD_b}
\end{figure}

Let's move to the first step of the charged-particle evaporation process: pure neutron removal. This process occurs predominantly in peripheral collisions, whose cross sections $\sigma_{-xn}$ over $\sigma_\text{cc}$ can be estimated using a simple geometric model,   
\begin{align}
    \frac{\sigma_{-xn}}{\sigma_\text{cc}} &\propto \frac{\pi b^2_\text{max}-\pi (b_\text{max}-\Delta b)^2}{\pi b^2_\text{max}} \nonumber \\
    &\propto 1-(1-\frac{\Delta b}{b_\text{max}})^2 \;.
    \label{eq:geo}
\end{align}
Here, $b_\text{max}\approx 1.2(A^{1/3}_\text{P}+A^{1/3}_\text{T})$ is the maximum impact parameter, and $\Delta b$ characterizes the interaction range of neutron removal. 
To validate this geometric picture, we simulate collisions of $^{14}$C with C and Pb targets using IQMD coupled with the GEMINI code, and extract the impact parameter ($b$) distributions for the charge-changing reaction channels and CPE, as shown in Fig.~\ref{fig:IQMD_b} (a). Although the charge-changing cross sections are significantly larger on lead than on carbon, the $b$ distributions of the evaporation for both targets share the same Full Width at Half Maximum (FWHM), 1.65 fm, implying a similar reaction overlap range $\Delta b$.
Substituting the value of $\Delta b$ into Eq.~\ref{eq:geo}, the results are shown in Fig.~\ref{fig:IQMD_b} (b). The neutron-removal fraction decreases systematically with increasing target mass, thereby reducing the relative evaporation probability. This explains why the stronger evaporation contribution is observed for low-$Z$ targets than for high-$Z$ ones.
Therefore, the target dependence of the indirect contribution primarily arises from the systematic decrease in the neutron-removal fraction as target size increases.

For $\sigma_\text{cc}$ measurements on high-$Z$ targets, one may wonder whether EMD can significantly affect $\sigma_\text{cc}$ as it does in $\sigma_\text{I}$. The answer hinges on the distinct nature of these observables. EMD plays an important role in neutron removal, strongly influencing $\sigma_\text{I}$ for neutron-halo nuclei or nuclei with loosely bound neutrons, owing to the dominance of $E1$ photons. An $E1$ giant resonance is excited only if the collision speed is much faster than the oscillation between protons and neutrons of the core. It is equivalent to high-energy $E1$ photons produced by the high-$Z$ target with a higher-energy beam of several GeV/nucleon~\cite{Bertu1988PR}.
In contrast, the electromagnetic contribution is fundamentally different and significantly smaller for $\sigma_\text{cc}$.
Proton knockout is generally suppressed in both stable and neutron-rich nuclei, as protons reside predominantly in the core. 
Although EMD may contribute to proton-halo nuclei, its cross section would be invisibly small at beam energies below approximately 1 GeV/nucleon~\cite{LiuJR2025PRC,Carlos2026Talk}. Moreover, the presence of the Coulomb barrier also hinders charged-particle evaporation. 
Taking $^{18}\text{O}$ on lead as an example,
the EMD contribution to $\sigma_{\text{cc}}$ at around 300 MeV/nucleon  
is estimated using the equivalent-photon method to be well below $1\%$~\cite{Carlos2026Talk, Bertu1988PR}. The same conclusion is obtained 
in the IQMD+GEMINI approach. For $^{107}\text{Ag}$ on lead at 240 MeV/nucleon, 
the total contribution from inelastic excitations from both electromagnetic and nuclear interactions to $\sigma_{\text{cc}}$ remains small ($\sim 0.2\%$). 
This suppression is physically understood by the hindrance of the high Coulomb barrier in heavy-heavy systems,
making high-$Z$ targets a better probe for extracting the proton radius. 

\label{section 4}
{\it Summary}---In summary, we measured 39 new charge-changing cross sections of 18 $p$-shell nuclides at around 240 MeV/nucleon on a number of targets. 
From these data, we report a unified description of the charge-changing cross-section scaling factor relative to the reaction model, elucidating the target-dependent behavior.  
The scaling factor decreases with increasing target mass and approaches 1 for the Pb target, reflecting the relative reduction in the neutron-removal cross sections for higher-$Z$ targets.   
We concluded that employing a heavy target, such as Pb, in charge-changing cross-section measurements enables the precise determination of proton radii for unstable isotopes, unlike in cases using low-$Z$ targets, where an empirical correction is essential. 
The robust scaling with separation energies is in principle applicable to all isotopes, provided that CPE remains the second dominant process. Further experimental data for nuclei approaching the drip lines are highly desirable to explore the generality of this scaling behavior.

{\it Acknowledgments}---We thank C. A. Bertulani for sharing the EMD calculations, and D.-Y. Pang for discussions on the reaction model. This work was partly supported by the National Natural Science Foundation of China (Grant Nos. 12325506,  12550008, and 12135004), the ``111 Center'' (B20065).

\bibliographystyle{apsrev4-2}
\bibliography{apssamp}

\end{document}